\documentclass[a4paper]{jpconf}
\usepackage{graphicx}
\begin{document}
\title{Heavy mesons in a hadronic medium: \\ interaction and transport coefficients}

\author{J M Torres-Rincon$^1$, L M Abreu$^2$, D Cabrera$^3$,  O Romanets$^4$ and L Tolos$^{3,5}$}

\address{$^1$ Subatech, UMR 6457, IN2P3/CNRS, Universit\'e de Nantes, \'Ecole de Mines de Nantes, 4 rue
Alfred Kastler 44307, Nantes, France}
\address{$^2$ Instituto de F\'isica, Universidade Federal da Bahia, 40210-340, Salvador, BA, Brazil}
\address{$^3$ Frankfurt Institute for Advanced Studies. Johann Wolfgang Goethe University,
Ruth-Moufang-Strasse 1, 60438 Frankfurt am Main, Germany}
\address{$^4$ KVI, University of Groningen, Zernikelaan 25, 9747AA Groningen, The Netherlands}
\address{$^5$ Instituto de Ciencias del Espacio (IEEC/CSIC), Campus UAB Carrer de Can Magrans s/n,
08193 Cerdanyola del Valles, Spain}

\ead{torresri@subatech.in2p3.fr}

\begin{abstract}
We review the recent results of heavy meson diffusion in thermal hadronic matter. The interactions of $D$ and ${\bar B}$
mesons with other hadrons (light mesons and baryons) are extracted from effective field theories based
on chiral and heavy-quark symmetries. When these guiding principles are combined with exact unitarity, physical
values of the cross sections are obtained. These cross sections (which contain resonant contributions) are used
to calculate the drag and diffusion coefficients of heavy mesons immersed in a thermal and dense medium. The
transport coefficients are computed using a Fokker-Planck reduction of the Boltzmann equation.
\end{abstract}

\section{Introduction}

  In relativistic heavy-ion collisions (HICs) heavy-flavor dynamics is considered one of the cleanest processes carrying direct information from the quark-gluon plasma (QGP) phase. 
During the last years ---and thanks to the experimental heavy-flavor results from the Relativistic Heavy-Ion Collider (RHIC) and the Large Hadron Collider (LHC)--- a lot of effort has been done in the theoretical side to understand 
a series of observables and infer the properties of the deconfined medium. Of special interest are the elliptic flow $v_2$ and the nuclear modification factor $R_{AA}$ of $D$ mesons, and of electrons coming from the heavy hadron decays.
These observables are sensitive to the microscopic details of hot matter and the interaction among their components. In particular, the transport coefficients of heavy particles affect these quantities
(in a similar fashion in which $v_2$ of light systems depends on the value of $\eta/s$ and $\zeta/s$). For these reasons, interaction and transport coefficients of heavy particles are fundamental ingredients to interpret the
experimental results of heavy particles and get knowledge of matter at extreme conditions.

  The last stage of a HIC is described by the expansion and final decoupling of the hadronic phase. In this medium, the heavy-flavor dynamics is dominated by the propagation of charmed and bottomed mesons and
baryons. Therefore, a complete description of a HIC should contain the evolution of these states in an expanding medium, and consequently, information about the slowing down of the momentum carried by them.
The relative amount of momentum stopping and diffusion is measured by the drag force and diffusion coefficients, respectively. To compute these transport coefficients from a microscopical framework, 
it is necessary to define a sensible effective theory for the interaction between heavy hadrons and the particles of the thermal bath.

  In this work we review our recent results on the microscopical calculation of the heavy-meson transport coefficients in a hot and dense medium using an up-to-date hadronic interaction based on effective field theories (EFTs) and demanding exact
unitarity. In particular, we will present our description for the interaction of $D$ and $\bar{B}$ mesons with lighter hadronic states, as well as the transport coefficients as functions of temperature and baryochemical potential.

\section{Heavy-meson interaction}

The EFT for the $D$ and $\bar{B}$ mesons follow quite in parallel because the meson mass (hard scale) dominates over all other scales~\cite{Manohar:2000dt,Abreu:2011ic}. 
For this reason the heavy-quark spin symmetry (HQSS) is one the key ingredients in the construction of the effective Lagrangian. At leading order, the interactions of $D$, $D^*$, $\bar{B}$ and $\bar{B}^*$ mesons are 
formally the same. This fact is reminiscent of the spin-flavor symmetry of the heavy-quark effective theory~\cite{Manohar:2000dt}. However physical masses explicitly break this symmetry, leading to
different interactions. Due to the presence of the pseudo-Goldstone bosons ($\pi,K,{\bar K},\eta$) the chiral symmetry is also exploited in the EFT.

We construct the effective Lagrangians for both meson-meson and meson-baryon interactions. The former can be arranged in a power-counting scheme giving a 
systematic expansion for the effective Lagrangian~\cite{Guo:2009ct,Abreu:2011ic,Abreu:2012et}. We consider the chiral expansion up to next-to-leading order, whereas the scattering
amplitudes are expanded at lowest order in the (inverse) heavy-meson-mass expansion.
In the meson-baryon sector the Lagrangian is based on a $SU(6) \times$HQSS symmetry, kept at leading order. The resulting Lagrangian coincides with a Weinberg-Tomozawa contact
vertex~\cite{Romanets:2012hm,GarciaRecio:2012db,Garcia-Recio:2013gaa}. In both sectors, we account for all possible elastic and inelastic channels involving a heavy-meson and a lighter hadron.

\begin{figure}
 \begin{center}
  \includegraphics[width=7.5cm]{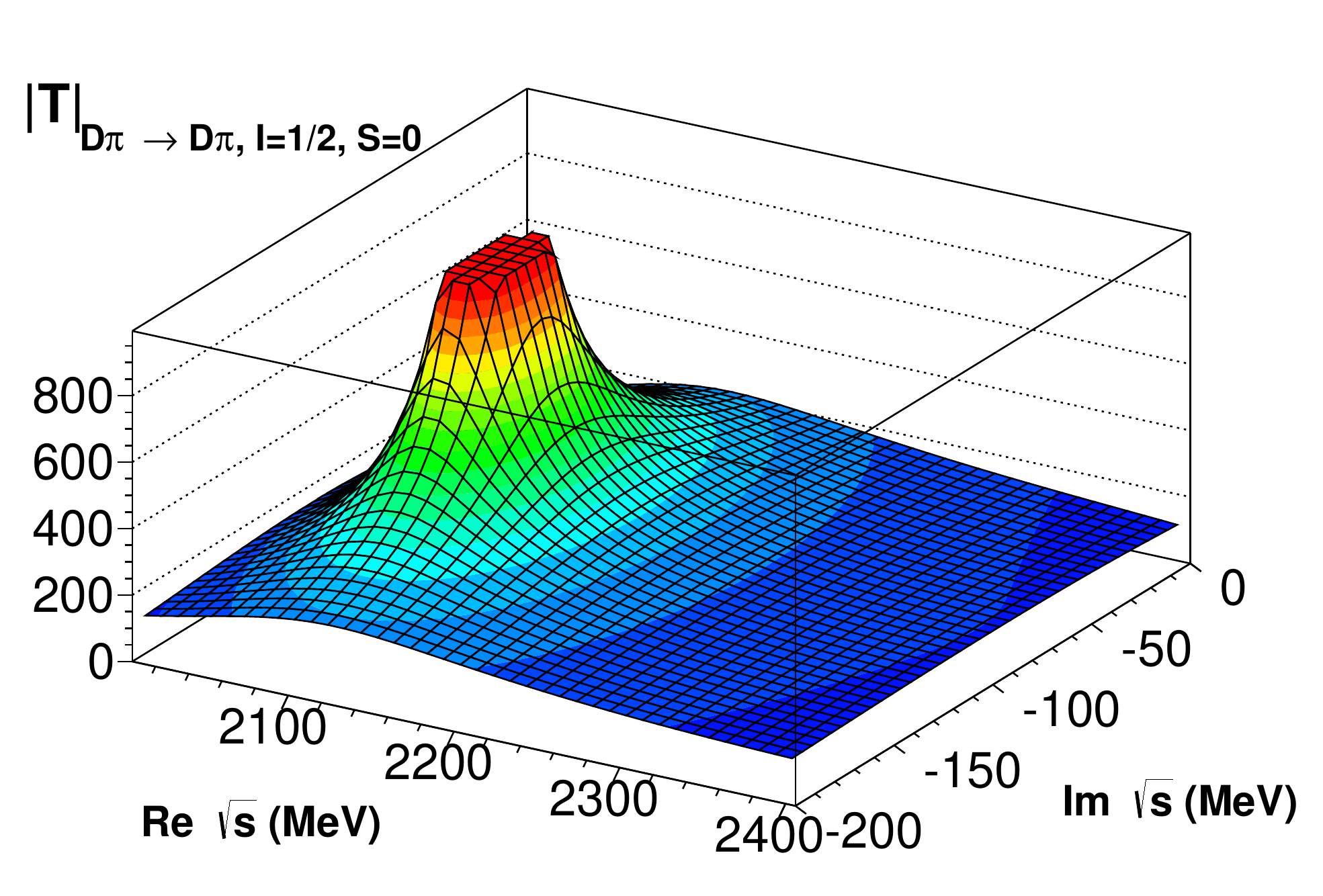}
  \includegraphics[width=7.5cm]{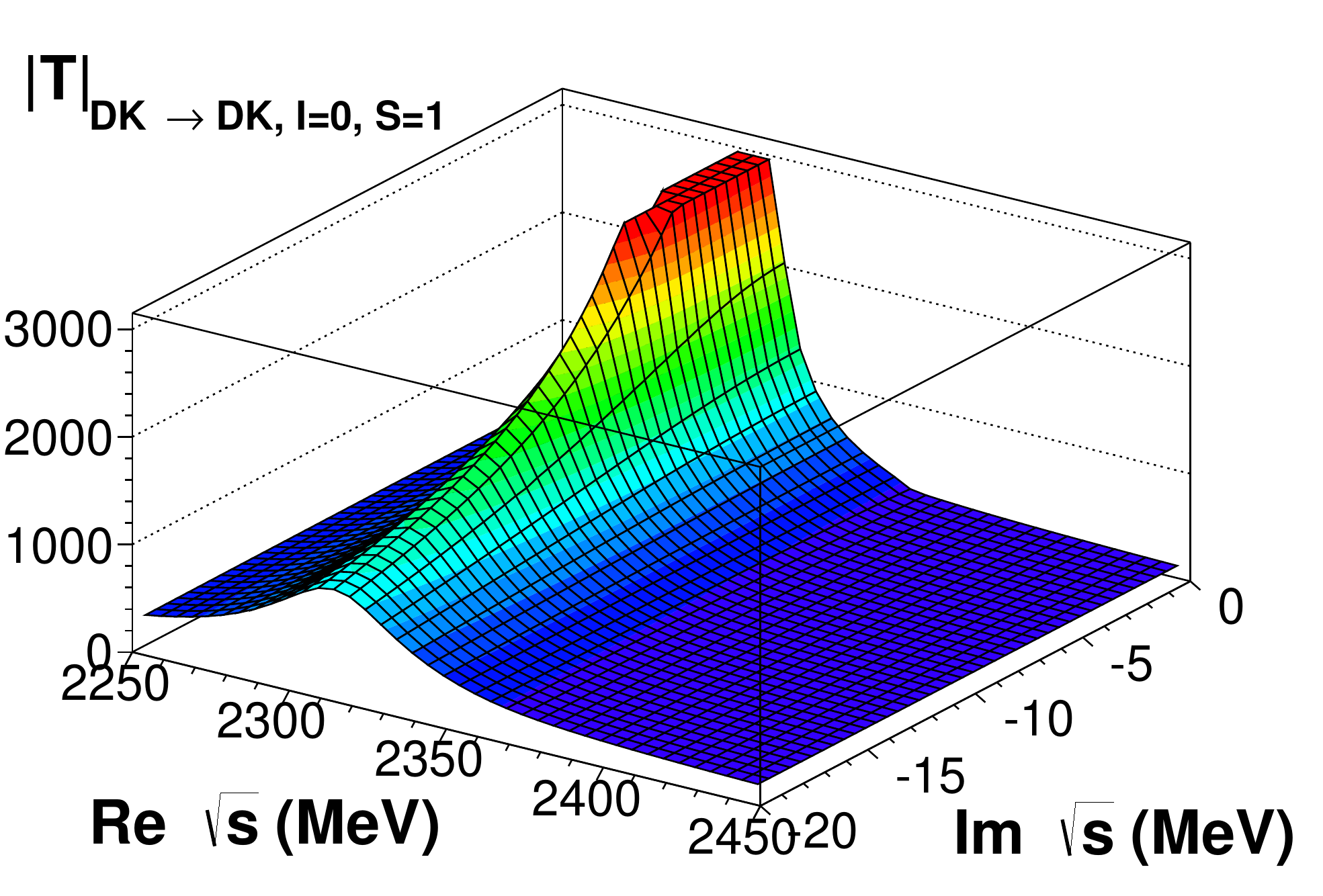}
 \end{center}
\caption{\label{fig:resonances} Many resonant states are generated when exact unitarity is imposed to the EFT scattering amplitudes. Left panel: Resonance in the $(I,S)=(1/2,0)$, $D\pi \rightarrow D\pi$ channel 
associated to the $D_0(2400)$ state. Right panel: Bound-state in the $(I,S)=(0,1)$, $DK \rightarrow DK$ channel associated to the $D_{s0}^{*}(2317)$ state.}
\end{figure}

We apply a unitarization scheme to ensure that the scattering-matrix elements satisfy the unitarity condition exactly. This is done by solving a Bethe-Salpeter
equation for the scattering amplitude in a coupled-channel basis~\cite{Tolos:2013kva,Torres-Rincon:2014ffa}. 
This method provides reliable interaction rates for $D, D^*, {\bar B}, {\bar B}^*$ mesons scattering off light hadrons: $\pi,K,{\bar K},\eta, N, \Delta$
(many other intermediate channels are also included e.g. $D_s \pi, \Lambda_c \eta...$ but they are not considered here as asymptotic states). 

In Fig.~\ref{fig:resonances} we show two channels in which resonant states are found in the scattering amplitude (note that the physical scattering amplitude is taken along the real-energy axis). 
A typical resonance ---with a finite decay width--- and a bound state ---without decay width--- are clearly seen in the left and right panels, respectively. It is important to notice that the EFT alone 
cannot give physical interactions without the application of a unitarization method, because the unitarization might give rise to resonances or bound states which produce strong modifications to the cross sections.

\section{Transport coefficients and application}

\begin{figure}
 \begin{center}
  \includegraphics[width=7.5cm]{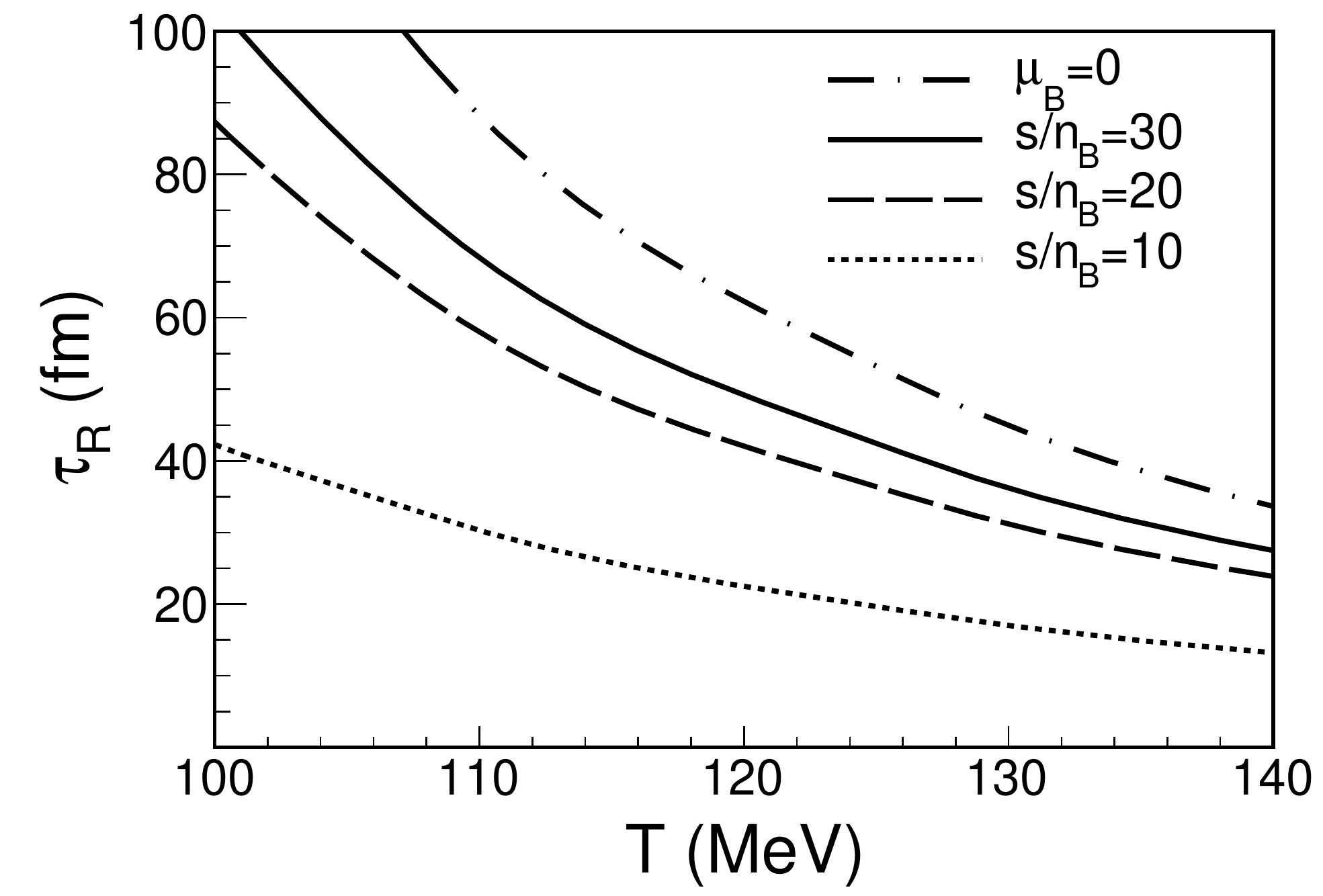}
  \includegraphics[width=7.5cm]{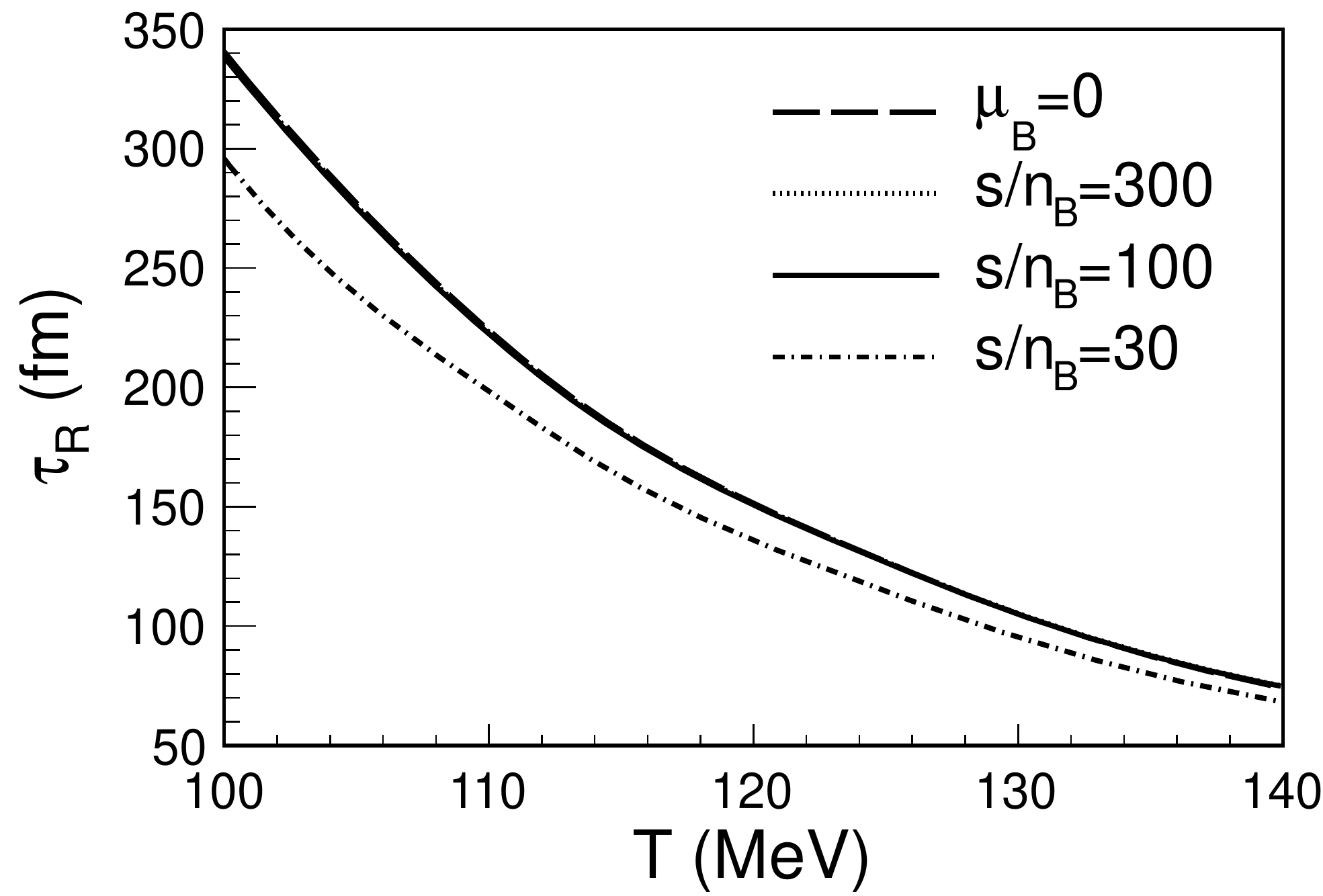}
 \end{center}
\caption{\label{fig:tau} Relaxation time for $D$ mesons (left panel) and ${\bar B}$ mesons	 (right panel) as a function of temperature for several isentropic trajectories.}
\end{figure}

\begin{figure}
 \begin{center}
  \includegraphics[width=7.5cm]{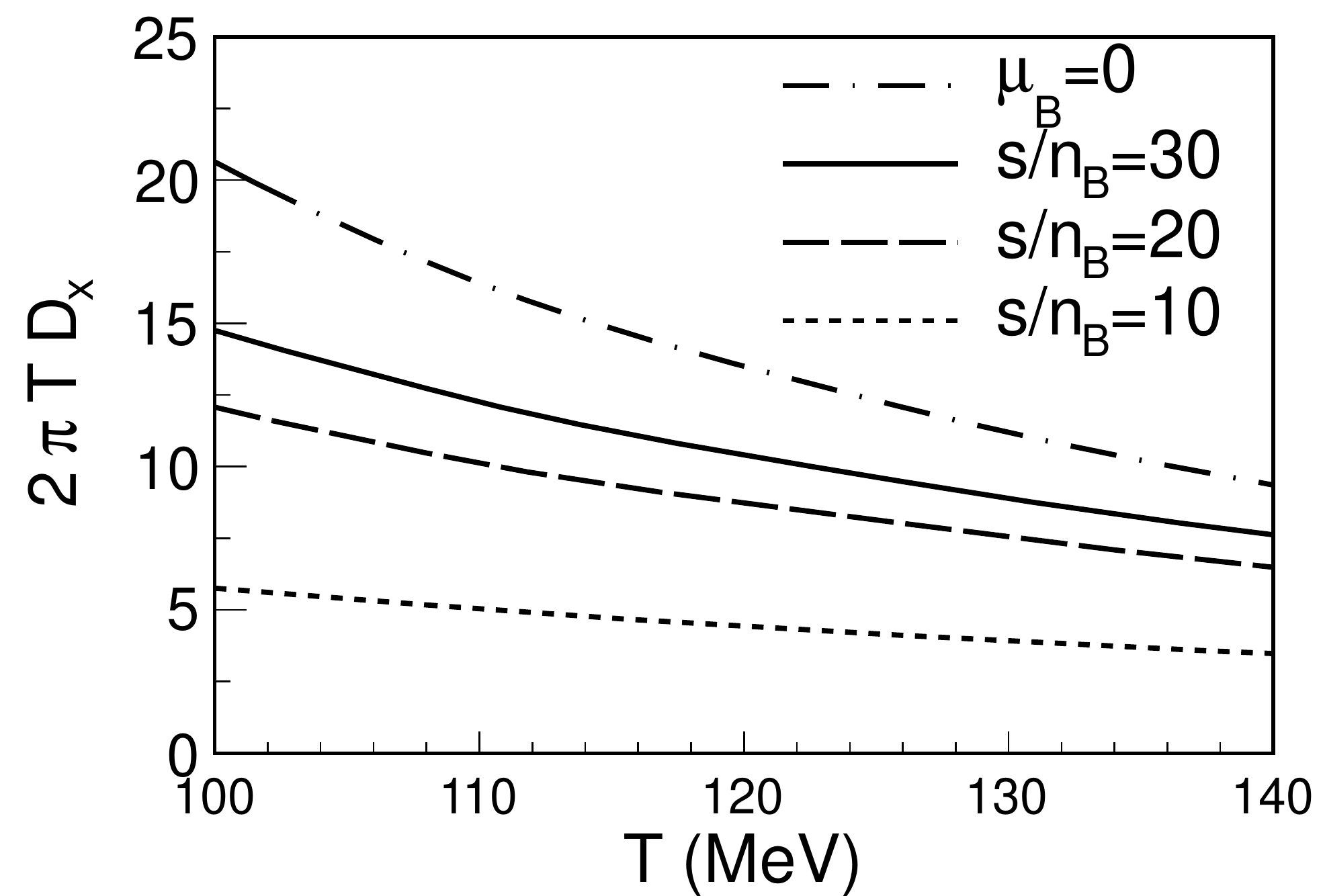}
  \includegraphics[width=7.5cm]{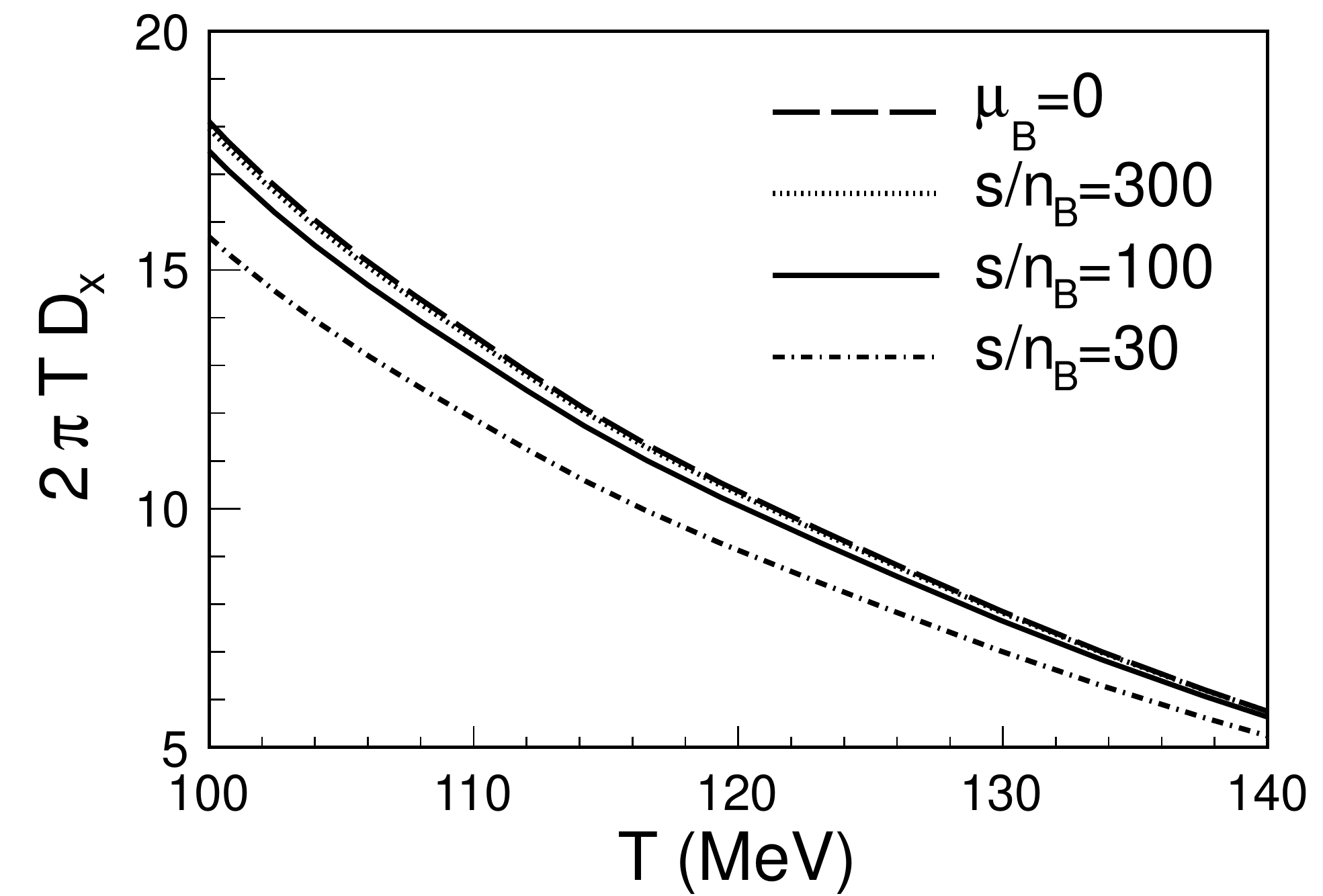}
 \end{center}
\caption{\label{fig:diff} Spatial diffusion coefficient for $D$ mesons (left panel) and ${\bar B}$ mesons (right panel) as a function of temperature and entropy per baryon.}
\end{figure}

The scattering amplitudes are now used to compute the transport coefficients of heavy mesons. The momentum distribution (initially out of equilibrium) evolves in time because of the rearrangement of 
momentum due to collisions. If we focus on the first two moments of the distribution ---i.e. average momentum and the distribution width---,
rather than the distribution itself, their time evolution is controlled by the drag force $F$ (that stops 
the heavy meson) and two diffusion coefficients $\Gamma_0, \Gamma_1$ (measuring the momentum broadening along the parallel and transverse directions of motion). These coefficients determine the Fokker-Planck equation governing the time evolution
of the distribution function. 
Derived quantities are the relaxation time of the average momentum $\tau_R=1/F$ and the spatial diffusion coefficient $D_x$ (which measures the spreading of heavy particles in the medium). 
We plot these coefficients for a heavy-meson momentum of $100$ MeV as functions of temperature and entropy per baryon in Figs.~\ref{fig:tau} and \ref{fig:diff}.

Based on estimates of non-relativistic kinetic theory and assuming a comparable interaction strength, the drag force should roughly scale with the inverse of the heavy-meson mass. For this reason the relaxation time
of the bottomed mesons is approximately $m_B/m_D \sim 3$ times larger than the one for charmed mesons (see Fig.~\ref{fig:tau}). On the other hand, the diffusion coefficient $D_x$ does not carry an explicit dependence 
on the heavy-meson mass. Therefore it should be approximately equal for $D$ and $\bar{B}$ mesons, as seen in Fig.~\ref{fig:diff} (different cross sections and subleading effects in the mass break this simple picture). 
Many other results can be found in our recent works~\cite{Tolos:2013kva,Torres-Rincon:2014ffa}.

It is interesting to mention that the diffusion coefficient has also been computed in the dynamical quasiparticle model for a quark-gluon plasma at finite temperature and chemical potential in Ref.~\cite{Berrehrah:2014tva}. The results
 ---with a sizable reduction with respect to perturbative quantum chromodynamics (QCD) calculations--- lie very close to our hadronic results around the transition temperature. This provides a nice agreement of the different models at both sides of the 
crossover region (also with lattice-QCD calculation around the transition temperature).
 
The results summarized here have been applied to a 2+1D MonteCarlo simulation of an ideal expanding plasma, showing that the hadronic interaction does not modify appreciably the $R_{AA}$ but contributes to the increase of the $v_2$~\cite{Ozvenchuk:2014rpa}.
Finally, in a dynamical quasiparticle model simulation the effect of hadronic interaction is a bit stronger, shifting the peak of $R_{AA}$ at
higher momenta and increasing $v_2$ up to a factor of 2~\cite{Song:2015sfa}. These examples show that the hadronic interaction presents a nonzero effect on the final observables, bringing the $R_{AA}$ and $v_2$ closer to the experimental results 
of RHIC experiments. We expect that the hadronic contribution will be more important for low-energy collisions like those at future 
Facility
for Antiproton and Ion Research (FAIR) or the Nuclotron based Ion Collider fAcility (NICA).

\ack

This work has been financed by grants FPA2010-16963, FPA2013-43425-P (Spain), BMBF (Germany) under project no.
05P12RFFCQ, EU Integrated Infrastructure Initiative Hadron Physics Project under Grant Agreement n. 227431 and Grant No. FP7-PEOPLE-2011-CIG under Contract No. PCIG09-GA-2011-291679.

JMTR acknowledges financial support from programme TOGETHER from R\'egion Pays de la Loire and the European I3-Hadron Physics programme. LMA thanks the Brazilian agencies CNPq (Project no. 308890/2014-0) and CAPES for financial support. 
DC acknowledges support from Centro Nacional de Fisica de Part\'iculas, Astropart\'iculas y Nuclear (Consolider - Ingenio 2010) and LT acknowledges support from the Ram\'on y Cajal research programme.

\end{document}